# Structure of chalcogenide glasses by neutron diffraction

G. J. Cuello[1,*], A. A. Piarristeguy[2], A. Fernández-Martínez[1], M. Fontana[3], A. Pradel[2]

[1] Institut Laue Langevin, 6 rue Jules Horowitz, BP 156, F-38042 Grenoble, France
[2] LPMC-CNRS, Université Montpellier II, 2 place E. Bataillon, F-34095 Montpellier, France
[3] Lab. de Sólidos Amorfos, Fac. de Ingeniería, UBA. Paseo Colón 850, (1063) Buenos Aires, Argentina.

___

**Abstract**

The purpose of this work is to study the change in the structure of the Ge-Se network upon doping with Ag. We report here a neutron diffraction study on two glasses of the system $Ag_x(Ge_{0.25}Se_{0.75})_{100-x}$ with different silver contents ($x$ = 15 and 25 at.%) and for two different temperatures (10 and 300 K). The total structure factor $S(Q)$ for the two samples has been measured by neutron diffraction using the two-axis diffractometer dedicated to structural studies of amorphous materials, D4, at the Institut Laue Langevin. We have derived the corresponding radial distribution functions for each sample and each temperature, which gives us an insight about the composition and temperature dependence of the correlation distances and coordination numbers in the short range. Our results are compatible with the presence of both $GeSe_{4/2}$ tetrahedra and Se-Se bonds. The Ag atoms are linked to Se in a triangular environment. Numerical simulations allowing the identification of the main peaks in the total pair correlation functions have complemented the neutron diffraction measurements.

___

## 1. Introduction

Silver-containing chalcogenide glasses have been extensively studied in the last decades. Of particular interest are their electrical conductivity that changes by several orders of magnitude upon silver doping. For example the conductivity of $Ag_x(Ge_{0.25}Se_{0.75})_{100-x}$ glasses suddenly changes by 7 orders of magnitude for $x \sim 10$ at.% [1-3]. It was recently shown by field effect-scanning electron microscopy and electrostatic field microscopy that the change occurred in these phase-separated glasses when the silver-rich phase starts connecting [4]. However the mechanism of diffusion throughout the glassy matrix is still not well understood. Structural investigation might help in getting a better insight to answer these questions.

This work is focused on the structure of two glasses within the fast ionic conduction region, *i.e.* $Ag_{15}(Ge_{0.25}Se_{0.75})_{85}$ and $Ag_{25}(Ge_{0.25}Se_{0.75})_{75}$, at two different temperatures. Neutron diffraction and MD simulations have been combined in order to have an insight of the short-range order in these glasses.

## 2. Experimental Details

### 2.1 Synthesis

Two $Ag_x(Ge_{0.25}Se_{0.75})_{100-x}$ samples with $x = 15$ and 25 at.% (hereafter named Ag15 and Ag25, respectively) were prepared with 4N elements (4N = 99.99%). In all cases, 12 g of materials were synthesised by placing the powdered elements in stoichiometric proportions in a cylindrical quartz ampoule. The ampoules were evacuated to a pressure of $\sim 10^{-5}$ mbar and sealed. After synthesis and homogenisation for 7 h at $T \sim 1200$ K in a furnace, the ampoules were quenched in an ice+water mixture to obtain the glassy materials.

### 2.2 Neutron characterisation

Neutron diffraction experiments were carried out on D4 instrument at the Institut Laue-Langevin (Grenoble, France) [5]. A monochromatic beam of 0.5 Å was produced by reflection in a Cu 220 crystal. Low- and room-temperature experiments were performed using a standard orange cryostat.

For the experiment, the samples were placed in cylindrical vanadium containers of 8 mm (outer diameter) and 0.1 mm thick. The beam size was 13 mm width and 50 mm height. Diffraction spectra for both samples were registered at 10 K and 300 K. The required ancillary measurements, *i.e.* vanadium rod (8 mm in diameter), empty cryostat/furnace, empty cell and boron carbide plate with the same dimensions than the sample, were also carried out.

Further to measurements, raw diffraction data underwent the usual corrections: the background was first subtracted and the standard absorption, multiple scattering and inelasticity corrections were then performed using the CORRECT code [6].

### 2.3 Modelling

MD simulations were carried out with the Density Functional Theory (DFT) based code, Vienna Ab-initio Simulation Package (VASP) [7, 8]. The Perdew-Burke-Enzerhof variation of the General Gradient Approximation to DFT was chosen to calculate the atomic ground state energy with VASP. The MD runs started from a randomly distributed model with the appropriate densities for the amorphous systems (0.03695 and 0.03901 atoms/Å$^3$ [9] for the Ag15 and Ag25, respectively). A first step of thermalisation was done at 2400 K using the Compass force-field, before starting the DFT calculations. A second thermalisation at 1400 K was performed with VASP, followed by a quenching down to 400 K during 7 ps, and a third thermalisation at 400 K during 4 ps. The MD time step was 2 fs.

## 3. Results

Figure 1 shows the normalised structure factors for the Ag15 sample at the two different temperatures (10 and 300 K). On the whole the shapes of the two curves are very similar with only some changes in the intensity of the main features. The same observation is true for the Ag25 glass. Figure 2 shows the normalised structure factors for the two glassy samples Ag15 and Ag25 at $T = 10$ K. Here also the main features are similar. However one can note a marked decrease in the first sharp diffraction peak at 1 Å$^{-1}$ when the silver amount increases as usual in oxide and chalcogenide glasses. Also while the second peak does not change from one sample to the other, the next ones shift to lower $Q$ values when the silver content increases. Similar trends for the structure factors were observed at 300K.

|      | $T$   | $N_{SeSe}$ | $N_{AgSe}$ | $N_{AgAg}$ | $r_{GeSe\ and\ SeSe}$ | $r_{AgSe}$ | $r_{AgAg}$ |
|------|-------|------------|------------|------------|------------------------|------------|------------|
| **Ag15** | 10 K  | 0.917 | 2.814 | 4.254 | 2.374 | 2.735 | 3.070 |
|      | 300 K | 0.901 | 2.647 | 4.665 | 2.381 | 2.741 | 3.066 |
| **Ag25** | 10 K  | 0.643 | 2.892 | 2.966 | 2.375 | 2.713 | 3.063 |
|      | 300 K | 0.687 | 2.647 | 2.971 | 2.384 | 2.724 | 3.058 |

Table 1. Coordination numbers and bond lengths (in Å) for $Ag_x(Ge_{0.25}Se_{0.75})_{100-x}$ glasses at $T = 10$ and 300 K.

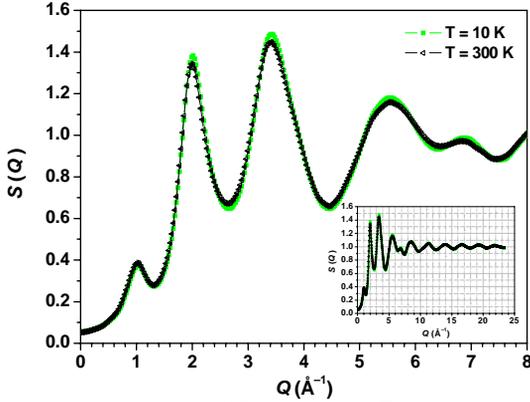

Fig. 1. Structure factors $S(Q)$, measured at $T = 10$ and 300 K for $Ag_{15}(Ge_{0.25}Se_{0.75})_{85}$ glasses.

Figure 3 depicts the simulated structure factor for the Ag15 glass along with that derived from neutron data. The simulations produced slightly different peak positions than experiments. Such a difference is due to the use of a soft pseudopotential for Se with a cut-off energy of 211 eV (the only one available at hand). It affected the coordinations in which Se was present by overestimating the correlation distances by about 4%. However, since the simulations were fairly well reproducing the trends in the experimental curves, the resulting atomistic model was used for calculating the partial pair correlation functions, $g_{ij}(r)$ for the two studied chalcogenide glasses. It was mainly used to interpret the experimental total pair correlation functions $g(r)$.

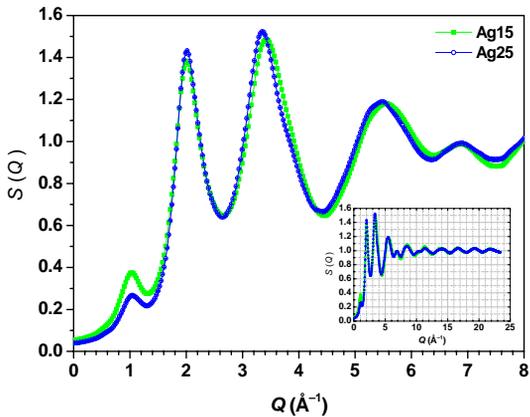

Fig 2 Structure factors $S(Q)$ for glasses $Ag_{15}(Ge_{0.25}Se_{0.75})_{85}$ and and $Ag_{25}(Ge_{0.25}Se_{0.75})_{75}$ at $T = 10$ K.

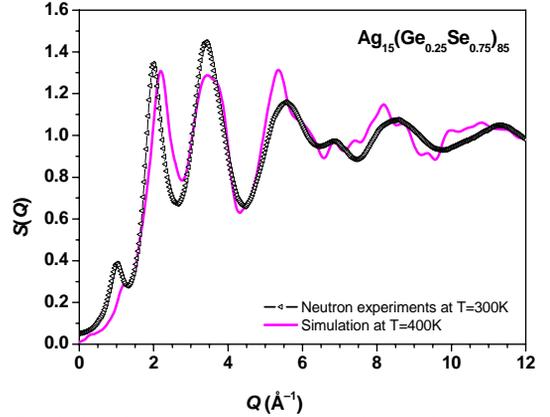

Fig. 3. Simulated structure factor $S(Q)$ for Ag15 sample at $T = 400$ K compared to neutron experiments at $T = 300$ K.

The total pair correlation functions were obtained by a Fourier transformation of the structure factors. The radial distribution function $RDF(r) = 4\pi r^2 \rho g(r)$, where $\rho$ is the density of the samples, was used for obtaining the correlation distances $r$ and coordination numbers $N$ in the system using the standard procedure [10]. In this purpose the different peaks were fitted using Gaussians. Simulation data helped us in identifying the peaks that could be attributed to the different coordination pairs. Figure 4 shows the radial distribution function for the Ag15 sample at 10 K and the Gaussian peaks corresponding to the correlations observed below 3.5 Å. According to simulation, the first peak had contributions from both Ge-Se and Se-Se pairs. A coordination number $N_{GeSe} = 4$ was assumed for Ge-Se which is in agreement with the generally accepted assumption that Ge is mainly present in a $GeSe_{4/2}$ tetrahedral environment in Ge-Se glasses with at.%Ge $\leq$ 33. Such an assumption helped us in calculating the $N_{Se-Se}$ coordination number. The next two peaks at about 2.38 Å and 3 Å were attributed to Ag-Se and Ag-Ag correlations respectively. Data obtained for both samples and both temperatures are given in Table 1.

## 4. Discussion

The structure factors $S(Q)$ and radial distribution functions RDF($r$) for both samples, Ag15 and Ag25, and both temperatures, 10 K and 300 K, showed the same main features. The obtained data are in good agreement with those obtained at 10K by Dejus *et al.* who studied an $Ag_{25}(Ge_{0.25}Se_{0.75})_{75}$ glass by isotopic substitution neutron diffraction [11]. The differences can be attributed to the different instrumental resolution for both experiments.

The first peak of the total pair correlation function (Figure 4) was attributed to the superposition of the correlation Ge-Se and Se-Se. Dejus *et al.* [11] did not include any Se-Se correlation in the analysis. Our choice was based upon both our simulation data and those by Tafen *et al.* [12] who studied two $Ag_{10}(Ge_{0.25}Se_{0.75})_{90}$ and $Ag_{15}(Ge_{0.25}Se_{0.75})_{85}$ glasses using the Firebold code. Owing to the ratio Ge/Se in both glasses, such correlations are expected if one accepts for the germanium the model of a $GeSe_{4/2}$ coordination polyhedron. A correlation distance for Ge-Se at 2.37-2.38 Å is in agreement with such a model since the average bond length of Ge and Se is 2.36 Å in the $GeSe_2$ crystal structure [13].

The next peak at about 2.71-2.73 Å corresponds to the Ag-Se correlations. A small shift in the peak position from 2.735 Å at $x = 15$ to 2.713 Å at $x = 25$ with increasing Ag was observed. A coordination number of ~ 2.8 Å was found for both glasses. The Ag-Se correlation distances are somewhat longer than those reported [11]. However they are in the range of the Ag-Se distances observed for two related crystalline compounds, *i.e.* $Ag_2Se$ and $Ag_8GeSe_6$. Both crystalline compounds present two phases, a low temperature one and a fast ion conducting high temperature one [14-17]. While the Ag atoms are disordered in the fast conducting phases at high temperatures, they are ordered in both low temperature forms. In this case, they have triangular and tetrahedral coordinations by Se with distances ranging from 2.62-2.86 Å. In the Ag15 and Ag25 glasses they clearly only retain the triangular coordination which is in agreement with Dejus's findings. According to these authors such a coordination might be a key in understanding the fast ion motion in these glasses. One can note that a similar coordination by S was also found for Ag in another fast ion conducting glass, *i.e.* $Ag_2GeS_3$ [18].

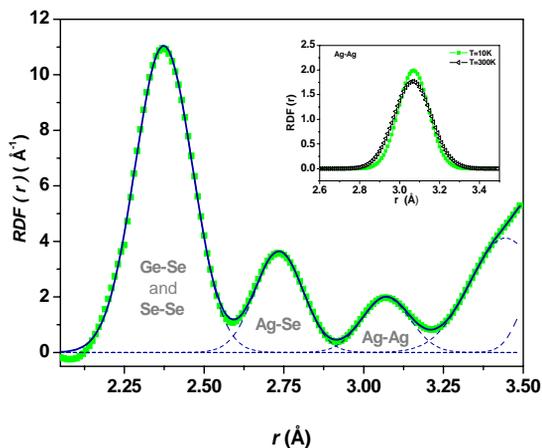

Fig. 4. Peak fitting in the radial distribution function RDF($r$) for Ag15 sample at $T = 10$ K . The Gaussian functions are depicted by dotted lines, the total fitted function by solid line and the measured function by the symbol.

The third peak in the total pair correlation function (Figure 4) was attributed to Ag-Ag correlations. For both samples a correlation distance of 3.06 Å was found. It is in agreement with both the experimental data from Dejus and the correlation distances that can be derived from the structure of the related crystalline phases, $Ag_2Se$ and $Ag_8GeSe_6$ (2.93-3.6 Å). However while the coordination number was found to be 4.2 for Ag15 sample it dropped down to 3.0 for the Ag-rich glass Ag25. The number for Ag25 disagrees with that found by Dejus *et al.* who rather proposed a coordination number of 3 as the one obtained for Ag15 [11]. However the authors reported a great difficulty in fitting the peak at 3 Å and had to proceed to a numerical calculation.

If we now look at the effect of the temperature on the structure of the glasses we mainly observe the expected changes due to increased thermal vibrations with a slight increase in length for Ag-Se and Ge-Se bonds. The main changes affect the peak at 3 Å attributed to the correlation Ag-Ag as shown in the inset of figure 4 for Ag15 sample. The peak decreases in intensity as a result of its broadening when the temperature increases. It can be explained by the increased diffusion of silver atoms at elevated temperature. Such an increased diffusion can also account for the decrease in coordination number of Ag by Se (refer to Table 1). Dejus *et al.* have also reported the broadening of peak at 3 Å. In their case it was a drastic change much larger than the one that we observe. Two facts can explain their results: they work at higher temperature, 100°C and 170°C and they report partial crystallisation of their sample with appearance of the Ag-rich argyrodite phase. Therefore the Ag-Ag correlations include those occurring in the argyrodite which is a superionic conductor with very high conductivity above room temperature [19].

## 5. Conclusions

The structure of ion conducting $Ag_{15}(Ge_{0.25}Se_{0.75})_{85}$ and $Ag_{25}(Ge_{0.25}Se_{0.75})_{75}$ glasses was studied by neutron diffraction and numerical simulations at 10 K and 300 K. Numerical simulations helped us in identifying the contribution of the different correlations to the RDF curves. It showed in particular the presence of both Se-Se and Ge-Se correlations in the first peak of the RDF that was not reported in previous experimental work. The results are consistent with a structure that contains both $GeSe_{4/2}$ tetrahedra and Se-Se bonds. The silver is bonded to Se in a triangular coordination. Changes in the Ag-Ag correlation peak at 3 Å with temperature is in agreement with increased diffusion of silver throughout the glasses, even though much less drastic changes than those reported previously for a similar study of Ag25 glass at higher temperatures were observed. Work is in progress to study the structure of Ag-Ge-Se glasses in the $x < 10$ at.% Ag range where the conductivity decreases by 7 orders of magnitude. The results will be shown in a forth-coming publication.

## 6. Acknowledgements

We are grateful to Pierre Palleau for his technical support during the diffraction experiments. We thank also Mark Johnson for the helpful advice and discussions.